\documentclass[12pt]{article}
\usepackage{graphicx}
\usepackage{subcaption}

\let\oldbibliography\thebibliography
\renewcommand{\thebibliography}[1]{%
  \oldbibliography{#1}%
  \setlength{\itemsep}{0pt}%
}

\usepackage{titlesec}
\titlespacing*{\subsection}
{0pt}{1.5ex plus 1ex minus .1ex}{1.5ex plus .1ex}

\newcommand\pubnumber{NuPhys2015-Coplowe}
\newcommand\pubdate{\today}

\def\oxford{Department of Physics\\
University of Oxford\\ Oxford\\ United Kingdom}

\def\email{\footnote{david.coplowe@physics.ox.ac.uk}}

\def\Title#1{\begin{center} {\Large #1 } \end{center}}
\def\Author#1{\begin{center}{ \sc #1} \end{center}}
\def\Address#1{\begin{center}{ \it #1} \end{center}}

\newcommand\pubblock{\rightline{\begin{tabular}{l} \pubnumber\\
         \pubdate  \end{tabular}}}
\newenvironment{Abstract}{\begin{quotation}  }{\end{quotation}}
\newenvironment{Presented}{\begin{quotation} \begin{center} 
             PRESENTED AT\end{center}\bigskip 
      \begin{center}\begin{large}}{\end{large}\end{center} \end{quotation}}

\def\beq{\begin{equation}}
\def\eeq#1{\label{#1}\end{equation}}
\def\eeqn{\end{equation}}

\def\beqa{\begin{eqnarray}}
\def\eeqa#1{\label{#1}\end{eqnarray}}
\def\eeqan{\end{eqnarray}}

\let\bar=\overbar

\def\Dslash{\not{\hbox{\kern-4pt $D$}}}
\def\dslash{\not{\hbox{\kern-2pt $\del$}}}

\def\msb{{\bar{\ssstyle M \kern -1pt S}}}

\begin{document}
\begin{titlepage}
\pubblock

\vfill
\Title{Isolating neutrino interactions on hydrogen in composite nuclear targets using the T2K Near Detector}
\vfill
\Author{David Coplowe\email\\on behalf of the T2K Collaboration}
\Address{\oxford}
\vfill
\begin{Abstract}
An analysis technique for isolating neutrino interactions on hydrogen, from a target containing a mixture of different nuclei, would provide numerous benefits. Namely, hydrogen is free of nuclear effects and enables better reconstruction of the neutrino energy spectra; key for neutrino oscillation experiments. Presented using Monte Carlo simulations of the ND280 near detector is the status of such a measurement on $\nu$-H resonance production by the T2K experiment.
\end{Abstract}
\vfill
\begin{Presented}
NuPhys2015, Prospects in Neutrino Physics\\
Barbican Centre, London, UK,  December 16--18, 2015
\end{Presented}
\vfill
\end{titlepage}
\def\thefootnote{\fnsymbol{footnote}}
\setcounter{footnote}{0}

\subsection*{Introduction}

A precise understanding of neutrino-nucleus interactions is paramount for current
~and future
~long-baseline neutrino oscillation experiments using sub to few GeV beam energies. In this energy regime, neutrino interaction uncertainties are dominated by our knowledge of nuclear dynamics. Isolating neutrino interactions on hydrogen enables the removal of such uncertainties providing concise measurements of the primary neutrino-nucleon interaction, and better neutrino energy reconstruction. Although such measurements have not been available experimentally for over three decades due to safety concerns, the advancement of new experimental techniques now make it possible to isolate interactions on hydrogen~\cite{Phys.RevD.92.5.051302.2015}.

Following the technique outlined in~\cite{Phys.RevD.92.5.051302.2015}, we highlight the status of such a method at the Tokai to Kamioka (T2K) near detector, ND280. The results are presented on Monte Carlo (MC) simulated data which uses the NEUT~\cite{NEUT} event generator and T2K neutrino flux~\cite{FLUX} to model neutrino interactions occurring in the ND280 detector equivalent to an exposure of $\sim7\times10^{21}$ POT. Defining the signal as a proton, positively charged pion, $\pi^+$, and muon, $\mu^-$, respectively in the final state, the hydrogen component is extracted using the double transverse momentum distribution --- a kinematic imbalance of the final state hadronic system transverse to the neutrino direction and muon momentum plane~\cite{Phys.RevD.92.5.051302.2015}.
 
\subsection*{ND280 Near Detector}
The ND280 detector~\cite{T2K} is situated 280 m downstream from the neutrino production target at an off-axis angle of 2.5$^\circ$, producing a peak beam energy of 600 MeV. The Inner Detector (ID) consists of an up-stream $\pi^0$-detector followed by a sandwich configuration of two hydrocarbon Fine-Grained Detectors (FGDs) nestled between three argon gas based Time Projection Detectors (TPCs) and a downstream Electromagnetic Calorimeter (ECal). The inner detector is surrounded by ECals providing a complementary measurement of energy from showers created by charged tracks within the ID and for neutral particle detection. These sub-detectors are situated within the UA1 magnet producing a 0.2 T magnetic field and is instrumented with muon detectors used to identify tracks coming from within or outside of the detector. 

\subsection*{Event Reconstruction and Selection}
Resonance production interactions via the $\Delta^{++}$-resonance are selected using a cut based analysis. Events are required to contain three tracks originating from a common vertex within the fiducial volume of FGD1. The neutrino direction is reconstructed from vertex information and the mean decay point of the neutrino parent. The three charged tracks must enter the TPC1 and pass the quality criterion outlined in~\cite{PRD92.112003.2015} to ensure good particle identification (PID) and reliable momentum reconstruction. The respective sign of the charged tracks is determined by the TPC defining the highest momentum negative charged track as the muon candidate. The remaining tracks must be identified as positively charged from which PID is undertaken to determine their particle type. For use in calculating transverse variables, like $\delta p_\mathrm{TT}$, the neutrino direction is reconstructed from vertex information and the mean decay point of the neutrino parent.
 
Note here that no particle identification is undertaken on the negatively charged track as the neutrino beam is almost purely muon-like (99\%)~\cite{PRD89.092003.2014}. Defining the signal as resonance production with a proton, pion and muon in the final state and no consideration as to the target, the selection results in a purity of 54\% for 3081 events with a contamination from electron neutrino interaction of the expected $\mathcal{O}$(1\%). The respective efficiency is 2\% for such a signal. In future studies an improved efficiency will be achieved by redefining the signal to include phase space constraints based on ND280 tracking thresholds.

The effect of a 46\% contamination from non-resonance interactions in double transverse momentum imbalance can be seen in Figure~\ref{fig:dpTTint}. Such backgrounds which are dominated by multi-pion production and produce a flat, symmetric distribution centred around zero. These background therefore have little influence on the shape of the $\nu$-H resonance signal enabling good signal determination using statistical methods.

\subsection*{Double Transverse Momentum}
\begin{figure}
\centering
\begin{minipage}{.496\textwidth}
\captionsetup{justification=centering}
  \centering
  \includegraphics[width=1.\textwidth]{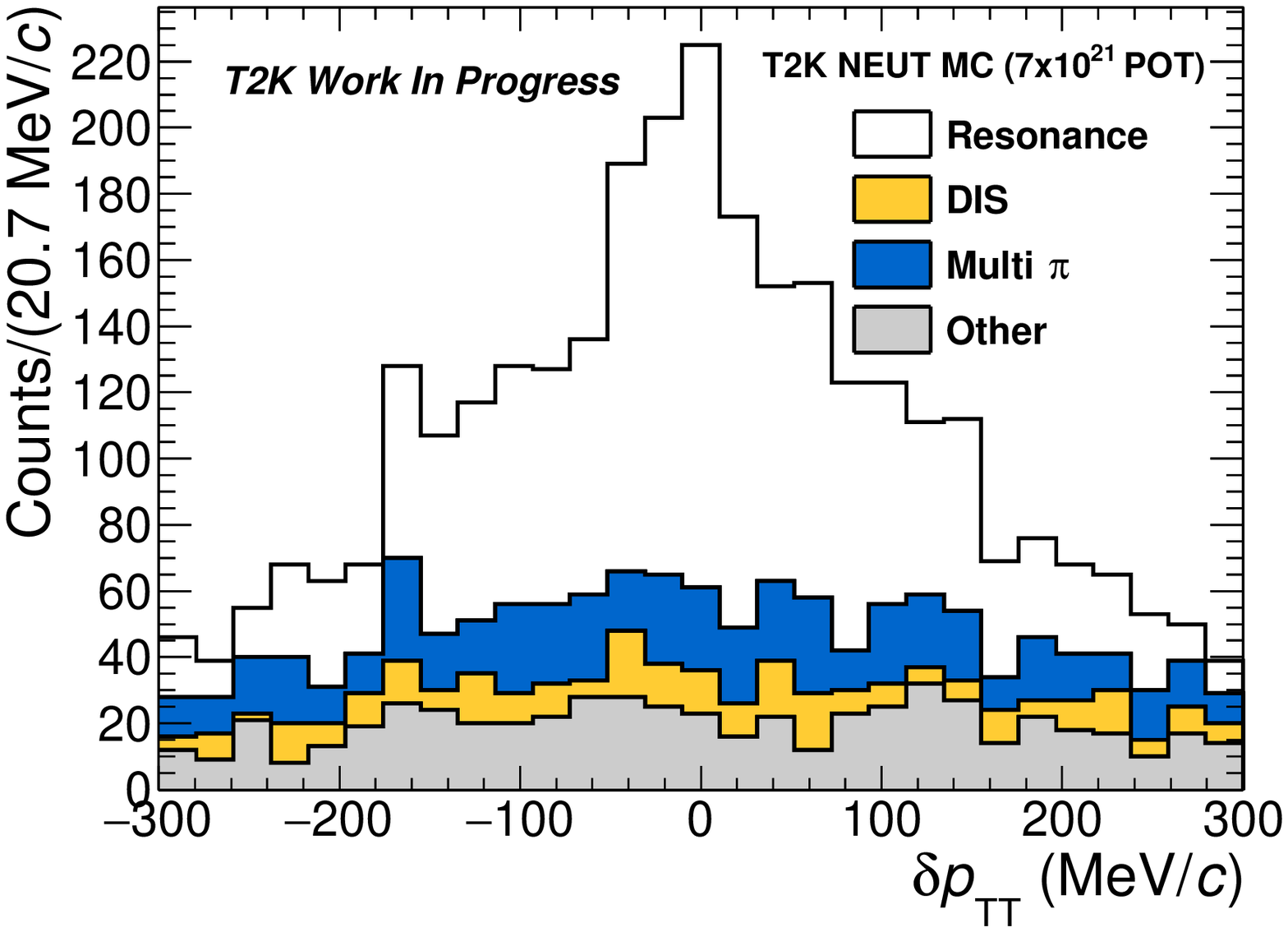}
  \captionof{figure}{Dominant interactions passing selection in reconstructed $\delta p_\mathrm{TT}$.\\ ~~~~~~~~~~~~~~}
\label{fig:dpTTint}
  \end{minipage}%
\begin{minipage}{.01\textwidth}\centering
\end{minipage}
\begin{minipage}{.496\textwidth}
\captionsetup{justification=centering}
  \centering
  \includegraphics[width=1.\textwidth]{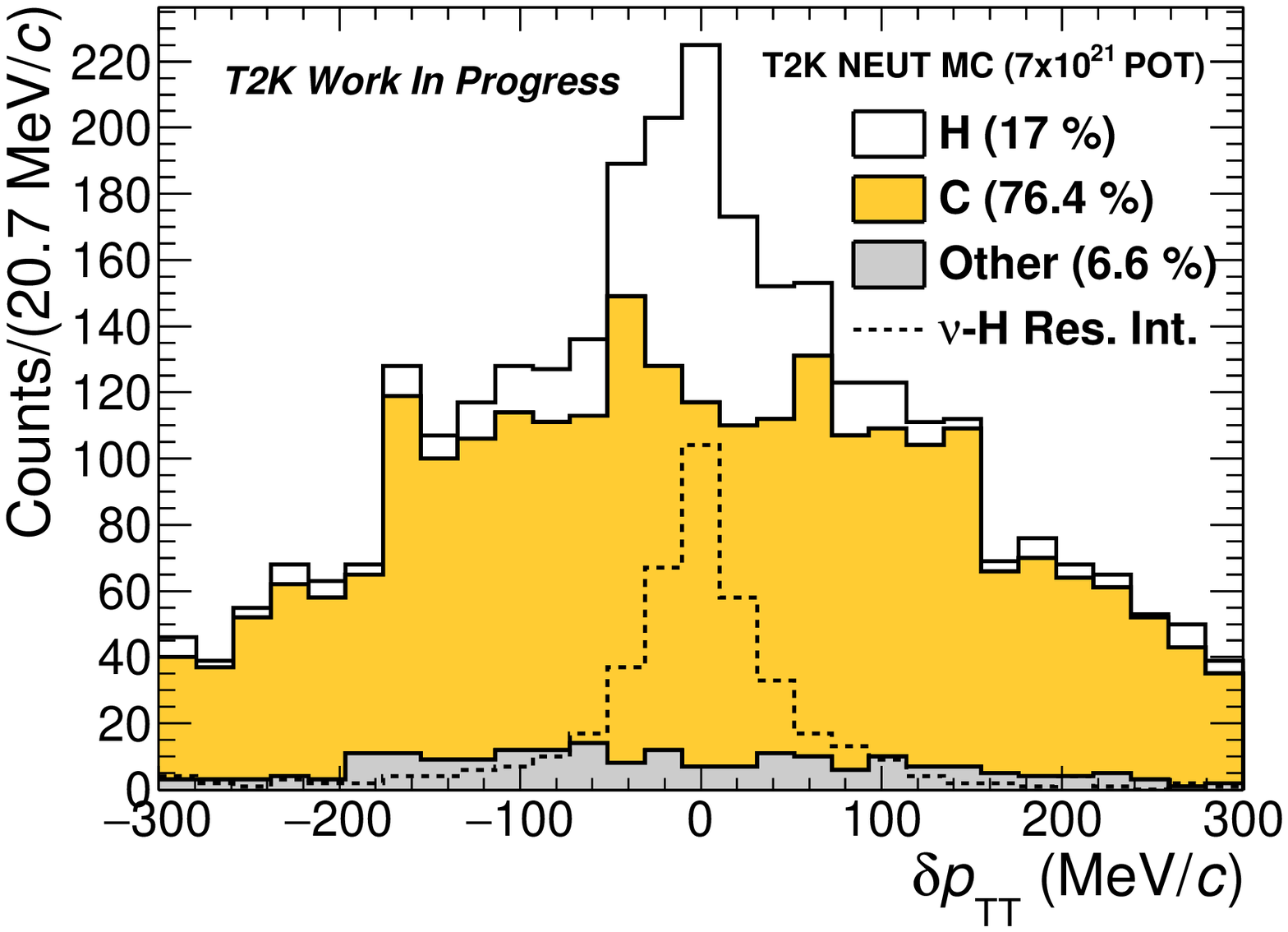}
  \captionof{figure}{Reconstructed $\delta p_\mathrm{TT}$ indicating relative target contributions. Dotted line emphasises signal.}
\label{fig:dpTTtarget}
\end{minipage}
\end{figure}

The double transverse momentum, $p^{p,\pi^+}_\mathrm{TT}$, is transverse to both the incoming and outgoing lepton and therefore absent of any lepton contribution. This results in any imbalance in $\delta p_\mathrm{TT} = p^{p}_\mathrm{TT}+p^{\pi^+}_\mathrm{TT}$ being a direct probe of nuclear effects~\cite{Phys.RevD.92.5.051302.2015}. Experimentally, detector smearing  broadens the double transverse momentum distribution. Figure~\ref{fig:dpTTtarget} highlights such an effect when considering the $\nu$-H resonance production component of $\delta p_\mathrm{TT}$. The dominant resonance, $\Delta(1232)$, from $\nu$-p(H) interactions produces a two body hadronic final state, $\pi^+$p, whose double transverse momentum component should balance, $p^\mathrm{p}_\mathrm{TT} = - p^{\pi^{+}}_\mathrm{TT}$ leaving $\delta p_\mathrm{TT} =0$. Such interactions are instead spread due to detector smearing. In contrast, neutrino interactions on nuclei with an atomic mass greater than one contain nucleons with some Fermi Motion (FM) causing an imbalance in $\delta p_\mathrm{TT}$. From Figure~\ref{fig:dpTTtarget} neutrino interactions are predominantly on carbon which has a FM $\mathcal{O}(217~\mathrm{MeV}/c)$
~and is seen to disperse events symmetrically around $\delta p_\mathrm{TT}=0$. The propagation of hadrons through the nuclear medium can result in Final State Interactions (FSI), changing particle types and final state kinematics. Monte Carlo studies of FSI at the neutrino interaction level indicate that FSI can be interpreted by a further spreading in $\delta p_\mathrm{TT}$. The smearing for the hydrogen target due to detector resolution is considerably smaller than the broadening for carbon due to additional nuclear effects and therefore enables extraction of the $\nu$-H resonance component.

\subsection*{Extracting Exclusive $\nu_\mu$-Hydrogen Events}
\begin{figure}
\centering
\begin{minipage}{.496\textwidth}
\captionsetup{justification=centering}
  \centering
  \includegraphics[width=1.\textwidth]{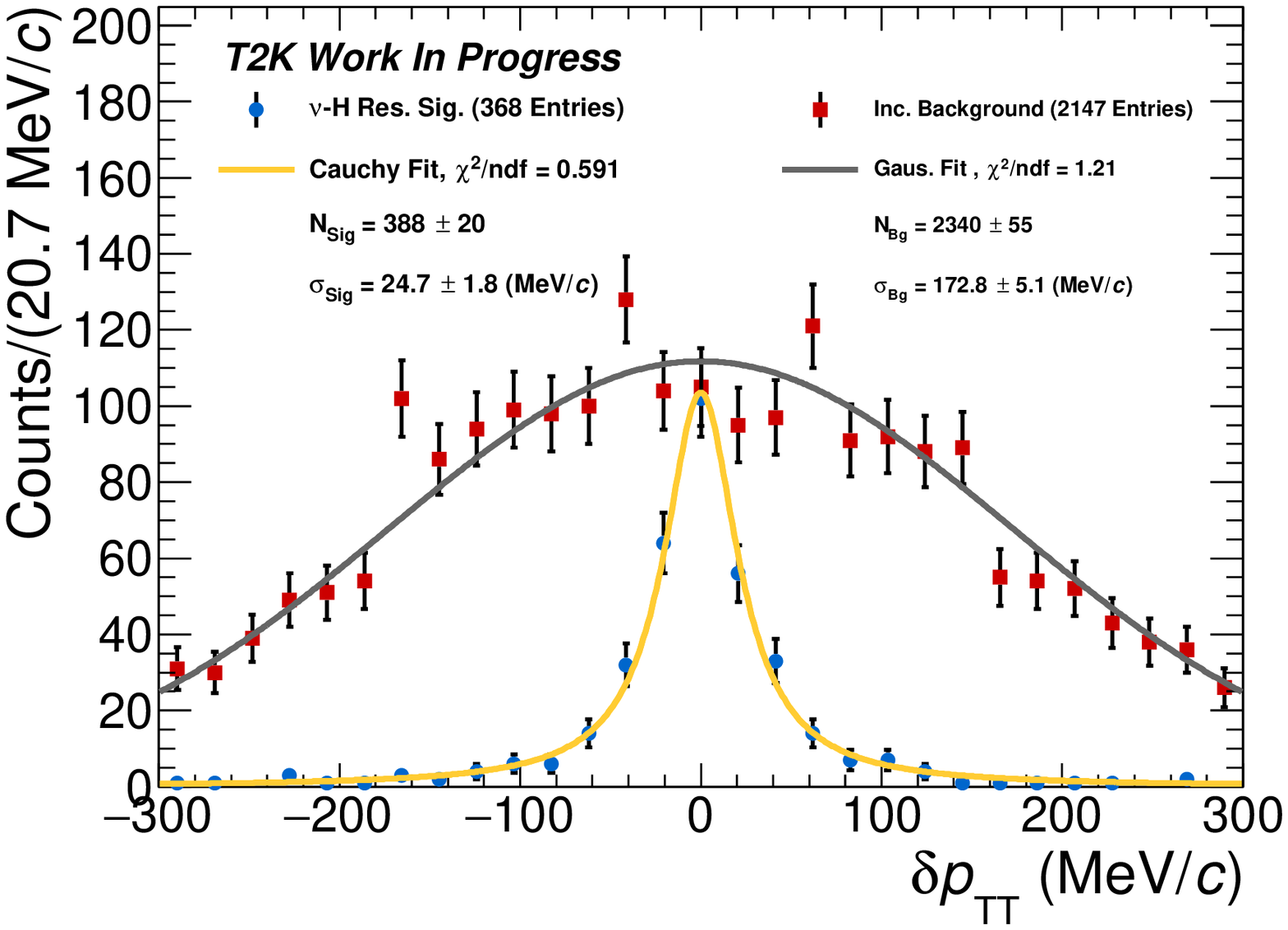}
  \captionof{figure}{Independent fits to Sig. and Bg respectively.}
\label{fig:indfit}
  \end{minipage}%
\begin{minipage}{.01\textwidth}\centering
\end{minipage}
\begin{minipage}{.496\textwidth}
\captionsetup{justification=centering}
  \centering
  \includegraphics[width=1.\textwidth]{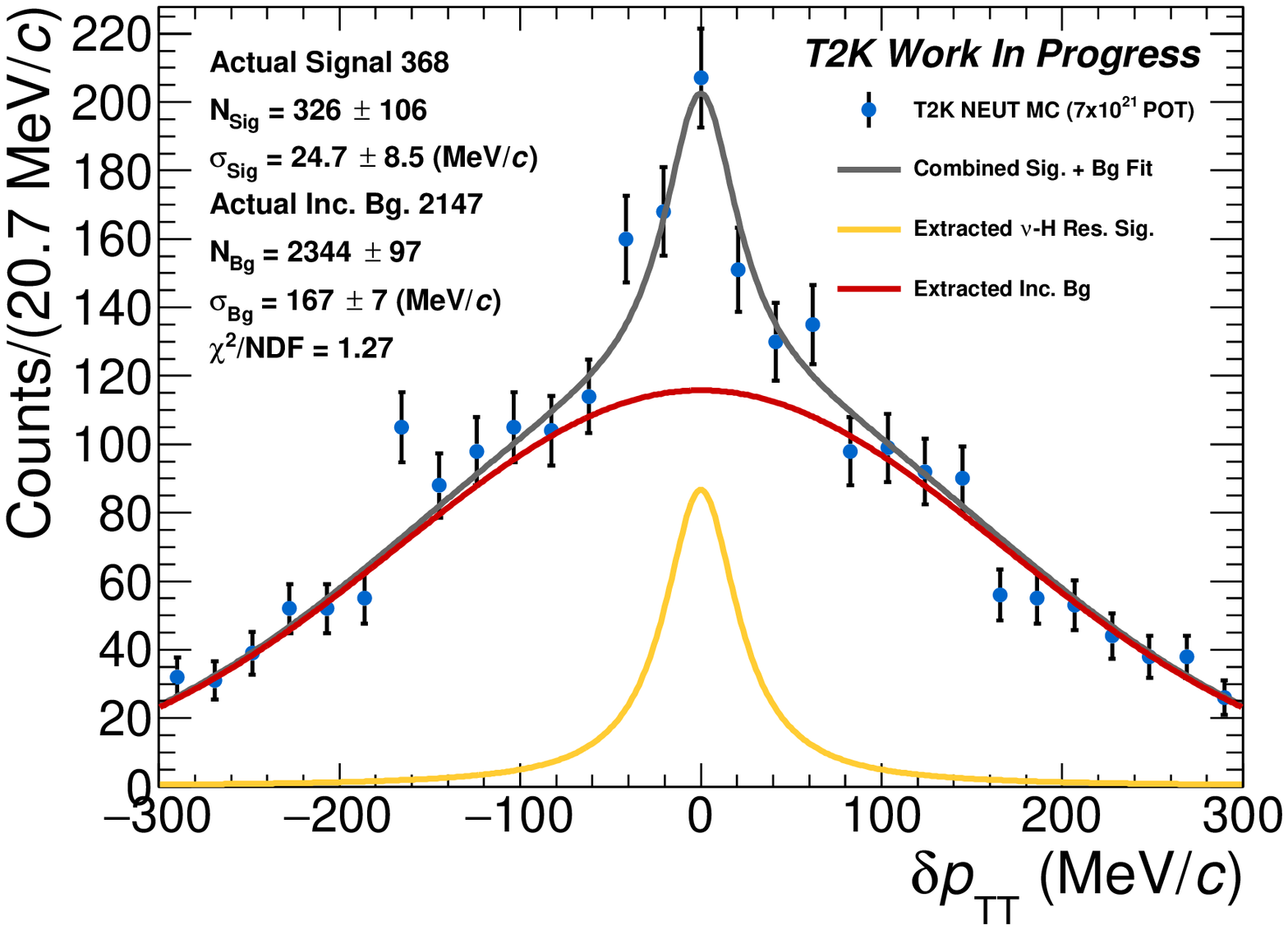}
  \captionof{figure}{Simultaneous fit extracting Sig. and Bg. event yields.}
\label{fig:comfit}
\end{minipage}
\end{figure}

At present, the hydrogen component is determined from  the $\delta p_\mathrm{TT}$ distribution by using a simultaneous log likelihood fit to extract the signal and background event yields. The exclusive $\mu,~\mathrm{p},~\pi^+$ signal is parametrised as a Cauchy function (motivated by smearing due to multiple scattering in the detector~\cite{Phys.RevD.92.5.051302.2015}) and accounts for detector smearing of $\nu$-H resonance events. A Gaussian distribution is used to model the background, while the systematic error due to the shape assumption will be investigated in the next stage of the analysis. Both functions are parametrised with three variables: a normalisation $N$, mean and width, $\sigma$ where the means are fixed at zero throughout the fitting procedures. Using the MC information on interaction type and target, the true signal and background are separated. The independent signal and background distributions are used to determine their respective widths for use as inputs to the simultaneous fit. The outcome of the independent fits can be seen in Figure~\ref{fig:indfit}. The fitted normalisation parameters $\mathrm{N}_\mathrm{Sig}$ and $\mathrm{N}_\mathrm{Bg}$ both overestimate the signal and background yields and are within one and four sigma of the actual yields. Using the widths $\sigma_\mathrm{Sig,Bg}$ as inputs to the simultaneous fit the outcome from fits to the same NEUT fake data used to determine the widths is given in Figure~\ref{fig:comfit}. It can be seen that there is little change in the widths determined by the fit when comparing it to the independent signal and background fits. The extracted background yield, $N_\mathrm{Bg}$, sees little change from that of the independent fit. This is not the case for the signal component which is now underestimated. These are however both within one and two sigma of their actual values.

\subsection*{Conclusion}
In this presentation, it has been shown that T2K's near detector, ND280, has the potential to perform a cross-section measurement exclusively on hydrogen --- the first in over three decades. The momentum and angular resolution of the TPC allows for good reconstruction of both proton and pion momenta and direction, allowing for the successful extraction of the hydrogen component from the nuclear background using the double transverse momentum variable~\cite{Phys.RevD.92.5.051302.2015}. The current focus is to moving towards a more sophisticated statistical method that accounts for detector and model systematics within the fit. This is especially important for correctly determining the background as theories used to model the nucleus are not well constrained. Given the current T2K dataset an absolute cross section measurement may be performed, enabling the extraction of model parameters free from nuclear effects. Such a measurement will improve our understanding of the dominant background for the single-ring-muon-like sample in T2K's muon neutrino disappearance measurements.

\end{document}